\documentclass[a4paper,fleqn,usenatbib,useAMS]{mnras}
\usepackage{graphicx}	% Including figure files
\usepackage{amsmath}	% Advanced maths commands
\usepackage{amssymb}	% Extra maths symbols
\usepackage{natbib}
\usepackage{graphicx}
\newcommand{\beq}{\begin{equation}}
\newcommand{\eeq}{\end{equation}}
\newcommand{\beqa}{\begin{eqnarray}}
\newcommand{\eeqa}{\end{eqnarray}}
\newcommand{\beqann}{\begin{eqnarray*}}
\newcommand{\eeqann}{\end{eqnarray*}}
\bibpunct{(}{)}{;}{a}{}{,}

\newcommand{\PhRvE}{Phys. Rev. E}

\begin{document}
\title[Propagation of Acoustic Waves in Inter-network]{Vertical Propagation of Acoustic Waves in the Solar Inter-Network as Observed by IRIS}

\author[P. Kayshap et al.]{
P.~Kayshap,$^{1}$\thanks{E-mail: virat.com@gmail.com}
K.~Murawski,$^{1}$
A.~K.~Srivastava,$^{2}$
Z.~E.~Musielak,$^{3,4}$
and B.~N.~Dwivedi$^{2}$
\\
% List of institutions
$^{1}$Group of Astrophysics, University of Maria Curie-Sk{\l}odowska, ul. Radziszewskiego 10, 20-031 Lublin, Poland\\
$^{2}$Department of Physics, Indian Institute of Technology (BHU), Varanasi-221005, India\\
$^{3}$Department of Physics, University of Texas at Arlington, Arlington, TX 76019, USA\\
$^{4}$Kiepenheuer Institut f\"ur Sonnenphysik, D-79104 Freiburg, Germany
}

\label{firstpage}
\pagerange{\pageref{firstpage}--\pageref{lastpage}}
\maketitle

\begin{abstract}

We investigate the Interface Region Imaging Spectrograph (IRIS) observations of the quiet-Sun (QS) to understand the propagation of acoustic 
waves in transition region (TR) from photosphere.  
%IRIS provides high resolution spectral observations in near ultraviolet (NUV) and far ultraviolet (FUV) wavelength ranges, which cover many photospheric and chromospheric/transition-region (TR) spectral lines.  
We selected a few IRIS spectral lines, which include the photospheric (Mn~{\sc i} 2801.25~{\AA}), chromospheric (Mg~{\sc ii} k 2796.35~{\AA}) and TR (C~{\sc ii} 1334.53~{\AA}), to investigate the acoustic wave propagation. 
%The analysis is performed using the wavelet cross-spectrum and coherence along with phase lags.  
The wavelet cross-spectrum reveals significant coherence (about 70\% locations) between photosphere and chromosphere. Few minutes oscillations (i.e., period range from 1.6 to 4.0 minutes) successfully propagate into chromosphere from photosphere, which is confirmed by dominance of positive phase lags. However, in higher period regime (i.e., greater than $\approx$ 4.5 minutes), the downward propagation dominates is evident by negative phase lags. The broad spectrum of waves (i.e., 2.5-6.0 minutes) propagates freely upwards from chromosphere to TR.  We find that only about 45\% locations (out of 70\%) show correlation between chromosphere and TR. Our results indicate that roots of 3 minutes oscillations observed within chromosphere/TR are located in photosphere. Observations also demonstrate that 5 minute oscillations propagate downward from chromosphere. \textbf{However, some locations within QS also show successful propagation of 5 minute oscillations as revealed by positive phase lags, which might be the result of magnetic field}. In addition, our results clearly show that a significant power, within period ranging from 2.5 to 6.0 minutes, of solar chromosphere is freely transmitted into TR triggering atmospheric oscillations. Theoretical implications of our observational results are discussed.
\end{abstract}
%} 
% -----------------------------------------------------------------------------------------
\begin{keywords}
Sun: oscillations -- Sun: photosphere -- Sun: chromosphere -- Sun: transition region
\end{keywords}
% -----------------------------------------------------------------------------------------
\section{Introduction}
% -----------------------------------------------------------------------------------------
Study of waves and oscillations in the different plasma settings of the solar atmosphere (i.e., umbra, penumbra, network, inter-network, 
loops, plumes, etc.) have significant importance as these structures are the prime candidates for transporting energy from the 
interior/photosphere to upper layers of the atmosphere \cite[e.g.,][and references cited therein]{Ofman1997,DeF1998,Sch1999,Moortel2002a, Moortel2002b,Rose2002,Centeno2006, Abhi2008, Sych2008, Centeno2009, Zaq2009,Jafer2017}.  Since the inter-networks (IN) regions are the places of very weak magnetic fields, they support acoustic waves.   These waves in the IN 
regions have been studied extensively to understand their nature as well as the wave propagation conditions in different layers 
of the solar atmosphere (\citealt{Lites1982,Lites1993,Carlsson1997,Judge2001,Wik2000,Bloomfield2004}).   The waves have 
attracted great attention as they consist efficient means of carrying energy between different layers of the solar atmosphere. \\

The acoustic cutoff frequency plays an important role in establishing the wave propagation conditions, and in finding regions in 
the solar atmosphere where the waves are strongly reflected.   The concept of acoustic cutoff  was originally introduced by \citealt{Lamb1909,Lamb1910}, who considered both an isothermal atmosphere as well as non-isothermal atmosphere with 
linear temperature profile (\citealt{Lamb1910,Lamb1932}).  Subsequently, Lamb's work was extended by a number of authors 
(e.g., \citealt{Moore1964, Sou1966,Fleck1993, Musielak2006, Fawzy2012,Routh2014}) who proposed different formulae for 
the cutoff period.  These analytical attempts were followed by a number of numerical simulations of acoustic waves (e.g., \citealt{Ulm1978,Carlsson1997,Fawzy2012}) with the aim of exploring the contribution of acoustic waves to atmospheric heating. \\

Recently, \citealt{Jim2006} and \citealt{Jim2011} reported variations of the cutoff frequency with the solar cycle and \citealt{Wisi2016} 
presented the first observational evidence for the existence of the acoustic cutoff in the solar atmosphere and showed its variations 
with the atmospheric height.   In addition, the same authors demonstrated that most analytical formulae previously obtained could 
not explain the observational data, which showed the cutoff increasing with height in the solar chromosphere.  \citealt{Murawski2016} approximately reproduced the observed variation of the cutoff period with atmospheric height by performing numerical simulations of impulsively generated acoustic waves.  However, in the numerical simulations performed by \citealt{Muraw2016}, acoustic waves were excited by a random driver, mimicking turbulence in the upper part of the convection zone, and it was found that wave periods follow the recent observational data (\citealt{Wisi2016}). 
The numerical simulations demonstrated as to how the solar atmosphere filters out the waves of low frequencies; clearly, these waves become evanescent and, therefore, they are absent in high layers of the solar atmosphere. High-frequency waves are free to propagate and reach higher layers of the solar atmosphere, \textbf{such as, \cite{Carlsson1992} have shown the conversion of higher frequencies (3 minute) into the chromosphere from lower frequencies (5 minute).}\\

In the context of wave propagation, \cite{Lites1979} proposed that oscillations below their frequency of 4 mHz become evanescent. 
However, higher frequencies (more than 4 mHz) freely propagate into the chromosphere from the photosphere.   Similar type of 
conclusions have been also reported in other papers (e.g., \citealt{Lites1982,Carlsson1992,Lites1993}).   In particular, \cite{Carlsson1992}
showed many aspects of the chromospheric inter-network oscillations, which are triggered by the photospheric motions.   They proposed 
that waves propagate upward from the photosphere.   Therefore, their amplitude grows and they become shocks in the upper chromosphere. 
In the paper, \cite{Carlsson1997} have shown that the shocks primarily result in the formation of IN bright grains. \cite{Hansteen2007} has 
reported that the waves above the acoustic cutoff frequency (5 mHz) in the IN regions can easily propagate up to the solar chromosphere. 
Recently, on the basis of numerical simulations, \cite{Murawski2016, Muraw2016} have also reported the propagation of high frequency 
acoustic waves into the chromosphere from the photosphere. \textbf{However, in case of magnetic field, the longer periods (i.e., periods more than 4 minuts) can easily propagate into the TR from photosphere as reported by \cite{Hegg2011}. Some observations also report the successful propagation of 5-minute in TR from photosphere within the regime of high magnetic fields (\citealt{DePon2003,DePon2005}).} Despite these achievements, it should be noted that the conditions for the \textbf{propagation of} acoustic wave \textbf{as well as waves in magnetic field} are still neither well-established nor understood.   Therefore, more observational results are needed to understand the origin, nature and behavior of acoustic waves in the solar atmosphere, and constraints on the propagation of these waves caused by the solar atmosphere.\\
 
In the present work, we have investigated the wave propagation in the QS region using IRIS spectroscopic observations. 
In Section~\ref{section:obs}, we present the observations and data analysis. Section~\ref{section:results} describes the observational 
results while discussion and conclusions are outlined in the last section.   
% -----------------------------------------------------------------------------------------
\section{Observations and Data Analysis}\label{section:obs}

IRIS provides high resolution observations (imaging as well as spectroscopic) for the broad range of altitudes in the solar atmosphere 
(i.e., the photosphere up to the corona; \citealt{DePon2014}). IRIS observed a quiet-Sun (QS region) on 16 November 2013 for approximately 
35 minutes from 07:33 UT to 08:08 UT. It captures the spectra in near ultraviolet (NUV) and far ultraviolet (FUV), which include various 
photospheric and chromospheric/TR lines (e.g., Mn~{\sc i} 2801.92~{\AA}, Mg~{\sc ii} k 2796.35~{\AA}, C~{\sc ii} 1334.53~{\AA}, 
etc.).   The observation is a sit-n-stare observation, which is appropriate to investigate the wave dynamics. In addition, this particular 
observation is located at the center of the solar disk (i.e., $\mu$$\sim$1.0). Therefore, different lines, which basically form at 
different heights, sample the same part of the solar atmosphere at different heights (no projection effect).
%%%%%%%%%%%%%Figure:1%%%%%%%%%%%%%%%%%%%%%%%%%%%%%%%%%%%%%%%%%%%%%%%%%%%%%%%%%%%%%%%%%%%%%%%%%%%%%%%%%%%%%%%%%%%
\begin{figure*}
\centering
\mbox{
\includegraphics[trim = 3.0cm 3.0cm 2.5cm 2.0cm, scale=1.1]{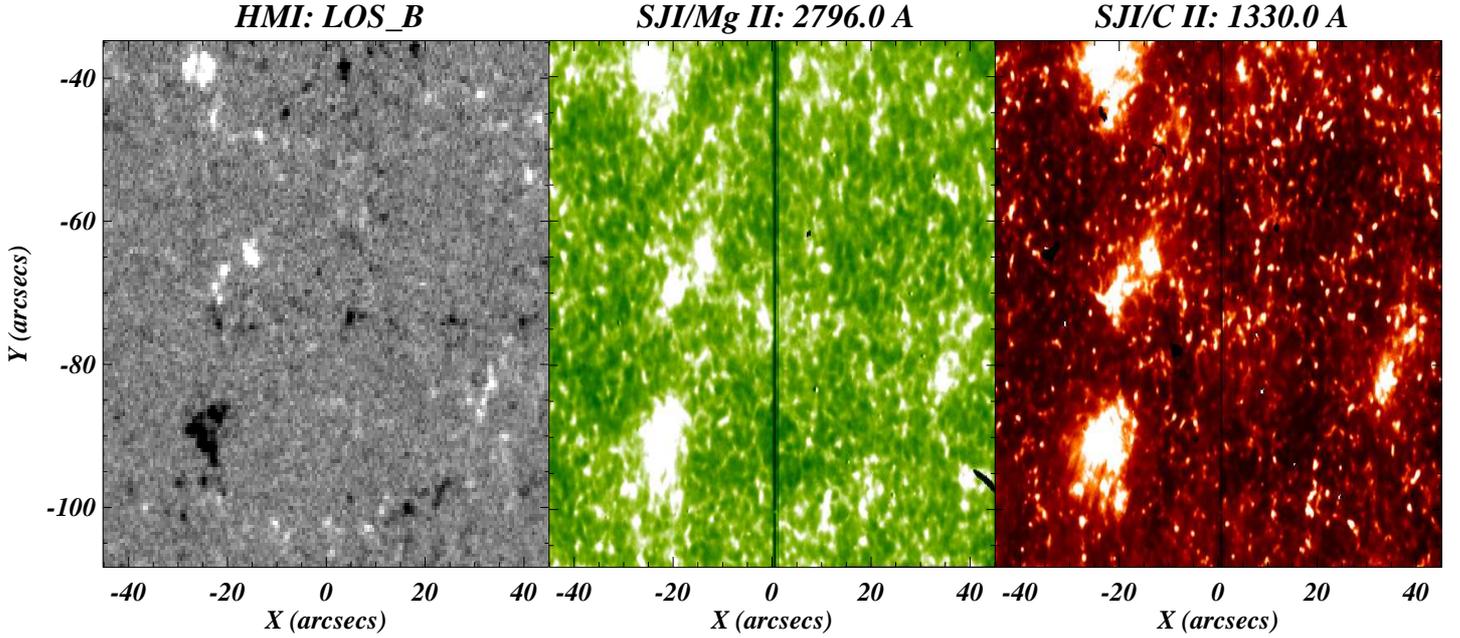}
}
\caption{\small LOS magnetogram (left panel), chromospheric (middle panel; IRIS/SJI Mg k 2796.0~{\AA}) and chromospheric view (right panel; 
IRIS/SJI C~{\sc ii} 1330~{\AA}) of the observed QS are presented. It is clearly visible that most of the region is permeated by very 
weak LOS magnetic field (left panel).}
\label{fig:ref_image}
\end{figure*}
%%%%%%%%%%%%%%%%%%%%%%%%%%%%%%%%%%%%%%%%%%%%%%%%%%%%%%%%%%%%%%%%%%%%%%%%%%%%%%%%%%%%%%%%%%%%%%%%%%%%%%%%%%%%%%%%%%%
Figure~\ref{fig:ref_image} shows the line-of-sight (LOS) magnetogram (left-panel), chromospheric (middle-panel) and TR (right-panel) 
images of the observed QS. The black line on IRIS/SJI Mg~{\sc ii} k 2796.0~{\AA} and C~{\sc ii} 1330.0~{\AA} filter shows slit location, 
which is used to take the spectra in various lines. The slit is located in the very quiet and low magnetic field area \textbf{(as revealed by magnetograms)} of this QS, which can be considered as the internetworks (IN). \textbf{However, some bright areas in the chromosphere/TR are also visible along the slit (cf., middle and right panel; Figure~\ref{fig:ref_image}), which may justify the presence of the significant magnetic field in those areas.}

The Mg~{\sc ii} k 2796.35~{\AA} line is an optically thick line, which covers 
a broad range of formation heights in the solar atmosphere.   Depending on the in-situ plasma conditions, the spectral profiles of 
Mg~{\sc ii} k~2796.35~{\AA} may show single, double or multiple peaks (\citealt{Leen2013}). However, most of the 
time Mg~{\sc ii} k 2796.35~{\AA} line shows double peaks in the QS.   As per \cite{Leen2013}, the blue/red peaks of the 
Mg~{\sc ii} k 2796.35~{\AA} line (i.e., k2v/k2r) originate from the middle chromosphere.   However, the central part of the 
Mg~{\sc ii} k 2796.35~{\AA} line (i.e., k3; dip region) forms in the upper solar chromosphere (just below the TR).\\
The Mn~{\sc i} 2801.25~{\AA} spectral line forms in the photosphere, which is an absorption line.   To estimate the peak intensity, 
Doppler velocity and line width, a single Gaussian has been fitted on the Mn~{\sc i} 2801.25~{\AA} line in the QS. 
We use inbuilt routine (i.e., iris$\_$get$\_$mg$\_$features$\_$lev2.pro; \citealt{Per2013}) to analyze the Mg~{\sc ii} k 2796.35~{\AA} 
line.   Using this routine, we estimate the Doppler velocities of k2v peaks of Mg~{\sc ii} k 2796.35~{\AA} lines. 
The C~{\sc ii} 1334.53~{\AA} spectral line samples the transition-region (TR), which is also an optically thick line. 
However, C~{\sc ii} 1334.53~{\AA} spectral line shows Gaussian profiles in this QS (most of the time), which can be fitted by 
single Gaussian to estimate the corresponding spectral properties. We have taken the photospheric (Mn~{\sc i} 2801.25~{\AA}), 
chromospheric (Mg~{\sc ii} k2v peak forms around 1500 km in middle chromosphere (\citealt{Vernazza1981}) \textbf{and \citealt{Leen2013} also show that Mg~{\sc ii} k2v forms around 1400 km in the middle chomosphere}) and TR heights (C~{\sc ii} 1334.53~{\AA}) for further analysis.   On the basis of selected lines, we cover the solar atmosphere 
from the photosphere up to TR to investigate the propagation of photospheric power into the higher layers.\\
%%%%%%%%%%%%%%%%%%%%%%%%%%%%%%%%%%%%%%%%%%%%%%%%%%%%%%%%%%%%%%%%%%%%%%%%%%%%%%%%%%%%%%%%%%%%%%%%%%%%%%%%%%%%%%%%%%%%%%%%%%%%%%
\begin{figure*}
\centering
\mbox{
\includegraphics[trim = 4.5cm 1.5cm 4.5cm 2.0cm, scale=1.1]{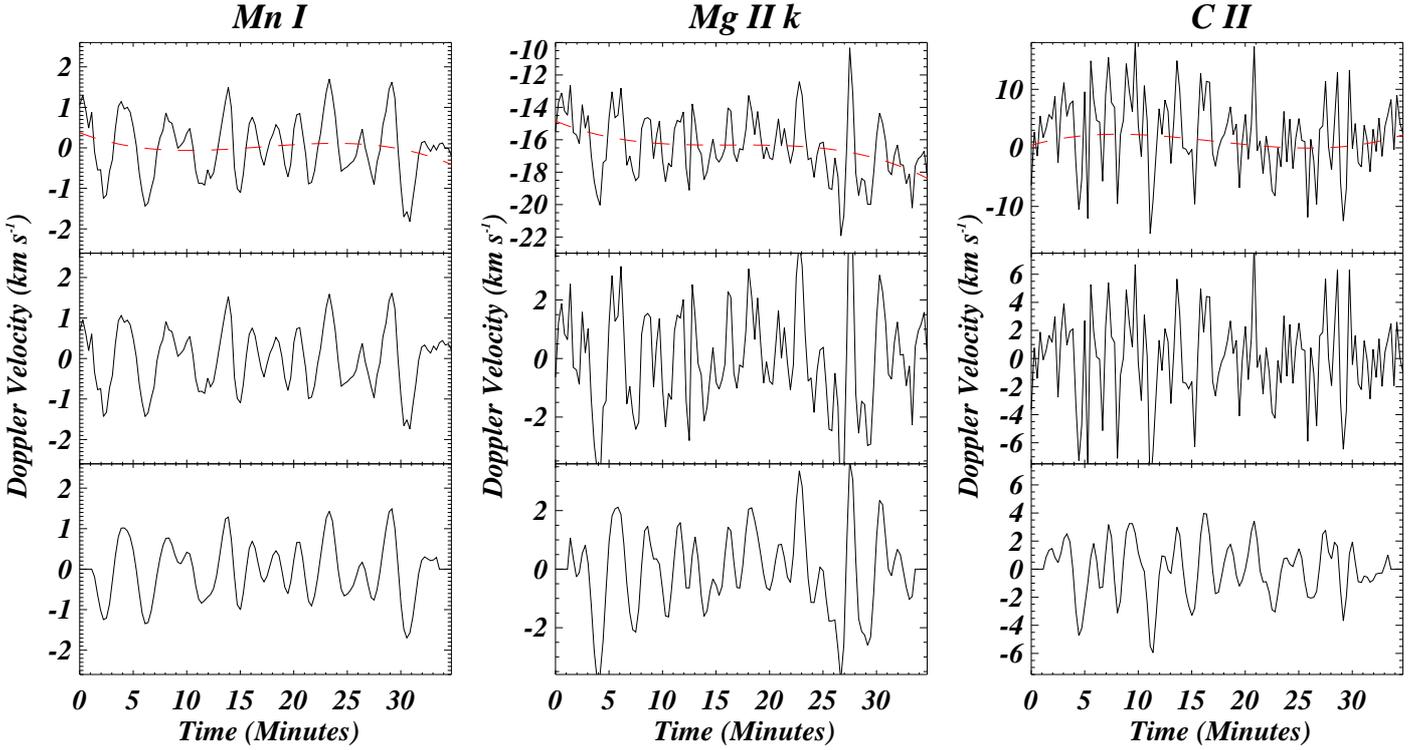}
}
\caption{\small Left column: top-panel shows the time-velocity series (solid line) of Mn~{\sc i} 2801.25~{\AA} along with its 
3$^{nd}$ order polynomial (dashed line), which is applied to remove the long-term trend. The detrended time series are shown in the 
middle-panel. We have applied a digital filter on the detrended time series to remove very high frequencies. In the bottom-panel, we 
have shown the time-series after removal the high frequencies from it. The similar results are also presented for Mg~{\sc ii} k2v 
(middle column) and C~{\sc ii} (right column).}
\label{fig:series_process}
\end{figure*}
%%%%%%%%%%%%%%%%%%%%%%%%%%%%%%%%%%%%%%%%%%%%%%%%%%%%%%%%%%%%%%%%%%%%%%%%%%%%%%%%%%%%%%%%%%%%%%%%%%%%%%%%%%%%%%%%%%%%%%%%%%%%%%%%%
The Doppler velocity-time series from these three lines are processed before the wavelet analysis.   First of all, all the time-series are 
detrended (i.e., removal of long-term trend using 3$^{rd}$ polynomial).   In the second step, we have removed the very high frequencies 
using a low-pass filter from all the time-series.   The left-column of Figure~\ref{fig:series_process} shows the original Doppler
velocity-time series (black curve; top-panel) along the 3$^{rd}$ polynomial (red-dashed line; top-panel) from Mn~{\sc i} 2801.25~{\AA}. 
The middle-panel of left column shows the detrended time-series of the same Mn~{\sc i} 2801.25~{\AA}.   Finally, the bottom-panel of 
left-column shows the time-series after removal of the high frequency components from it.   In the similar fashion, we 
have shown the processed time-series for Mg~{\sc ii} k2v (middle column) and C~{\sc ii} spectral liens (right-column). 
After processing these time series, we have applied the wavelet analysis (i.e., wavelet transform, cross-power and phase analysis).
%%%%%%%%%%%%%%%%%%%%%%%%%%%%%%%%%%%%%%%%%%%%%%%%%%%%%%%%%%%%%%%%%%%%%%%%%%%%%%%%%%%%%%%%%%%%%%%
% -----------------------------------------------------------------------------------------
\section{Observational Results}\label{section:results}
% -----------------------------------------------------------------------------------------
The wavelet analysis is an extremely useful tool for the simultaneous diagnosis of the power in time and frequency domains for the 
time-series.The wavelet transform is superior to the Fast Fourier Transform (FFT) due to its ability to diagnose the power in time and 
frequency, simultaneously. The wavelet analysis (i.e., wavelet transform, coherence and phase-lags) is suitable for searching transient 
oscillations and propagating waves through different layers of the solar atmosphere (\citealt{Bloomfield2004,McIntosh2004,Jess2007,
Jafer2017}). The wavelet transform is basically the convolution between the time-series and the $"$mother$"$ function. These are 
different types of inbuilt $"$mother$"$ functions (e.g., Morlet, Paul and Derivative of Gaussian (DOG)). For the present analysis, 
we have used $"$Morlet$"$ as mother function with a dimensionless frequency of $\omega_{0}$ = 6, which is suitable for investigating the 
propagation of waves with different range of frequencies (\citealt{Jafer2017}). The Morlet function is basically a complex wavelet, 
which is the resultant of plane wave modulated by the Gaussian function, that is 
%%%%%%%%%%%%%%%%%%%%%%%%%%%%%%%%%%%%%%%%%%%%%%%%%%%%%%%%%%%%%%%%%%%%%%
\begin{equation}
\label{eq:morlet}
{{\psi_{0}(\eta)}} = \pi^{-1/4} e^{i\omega_{0}\eta} e^{-\eta^2/2} \, 
\end{equation}
%%%%%%%%%%%%%%%%%%%%%%%%%%%%%%%%%%%%%%%%%%%%%%%%%%%%%%%%%%%%%%%%%%%%%%%%%%     
The wavelet transform provides a 2-D complex array for a time-series, which contains the distribution of power in frequency and time domain.
The power of the wavelet is defined as the square of absolute magnitude of 2-D complex array. The averaged wavelet power (i.e., over the 
time) can be considered as the global wavelet power, which is suitable to reveal the dominant oscillations for any particular time series.
%%%%%%%%%%%%%%%%%%%%%%%%%%%%%%%%%%%%%%%%%%%%%%%%%%%%%%%%%%%%%%%%%%%%%%%%%%%%%%%%%%%%%%%%%%%%%%%%%%%%%%%%%%%%%%%%%%%%%%%%%%%%%%
\begin{figure*}
\centering
\mbox{
\includegraphics[trim = 4.5cm 0.0cm 5.5cm 0.0cm, scale=1.2]{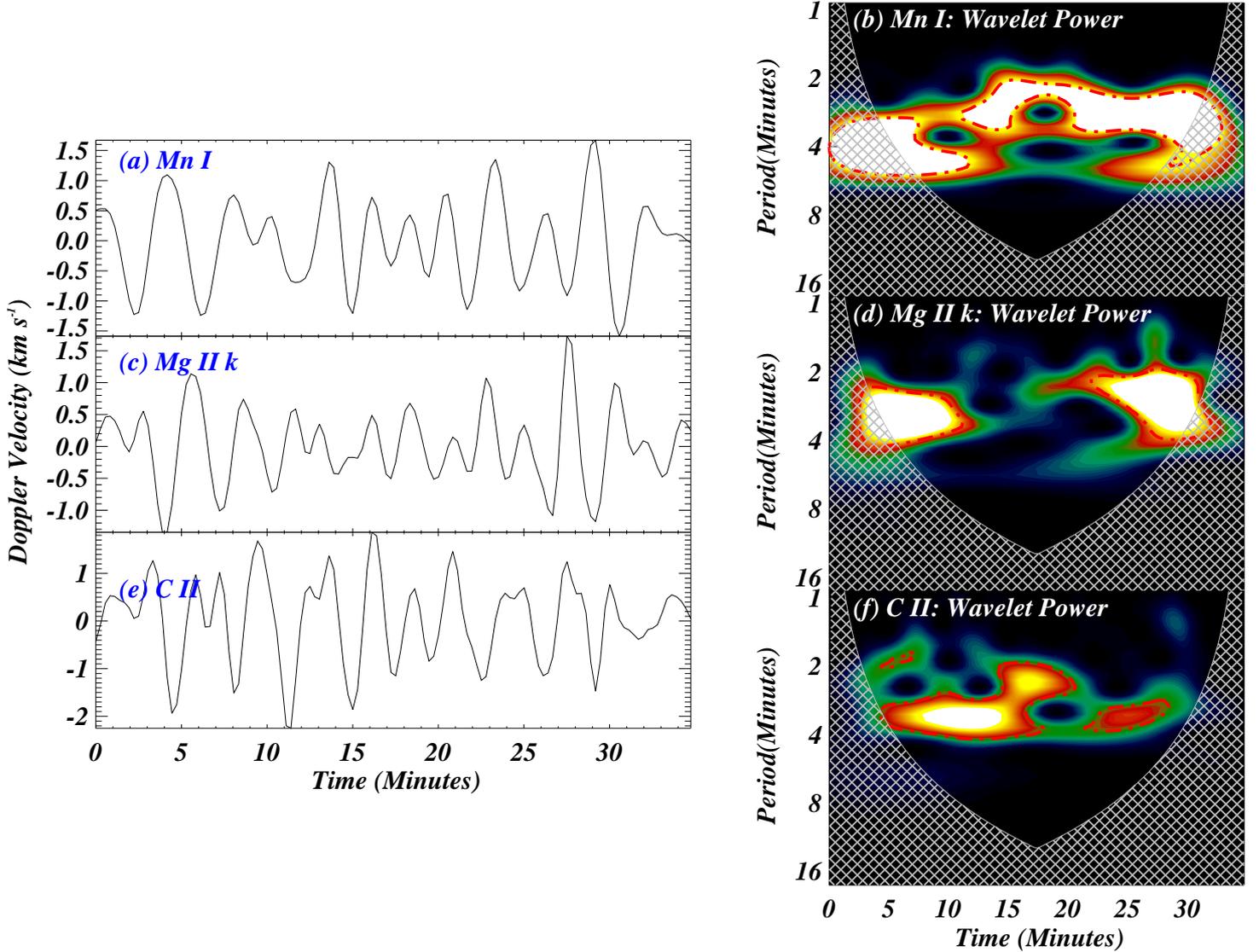}
}
\caption{\small The velocity-time series (panel a) and corresponding wavelet power map (panel b) of Mn~{\sc i} 2801.25~{\AA} line from one 
location. The significant power resides in between 2.0 to 6.0 minutes. Panel c and d shows the velocity-time series and wavelet power map 
from Mg~{\sc ii} k2v. In case of the chromosphere, the significant power lies in the period regime from 2.5 to 5.0 minutes. The velocity-time 
series and wavelet power map from the TR are shown in panel e and f. A significant power corresponds to 2.0-4.0 minutes for the TR. Red-dashed
lines on each wavelet map outlines the 95\% significance level and the gray hatched area outlines the COI.}
\label{fig:wavelet}
\end{figure*}
%%%%%%%%%%%%%%%%%%%%%%%%%%%%%%%%%%%%%%%%%%%%%%%%%%%%%%%%%%%%%%%%%%%%%%%%%%%%%%%%%%%%%%%%%%%%%%%%%%%%%%%%%%%%%%%%%%%%%%%%%%%%%%%%%
Figure~\ref{fig:wavelet} shows the Doppler velocity-time series (panel a) of Mn~{\sc i} 2801.25~{\AA} and corresponding wavelet power 
(panel b). The wavelet power of Mn~{\sc i} 2801.25~{\AA} shows that the  significant power lies inbetween 2.0 to 6.0 minutes. Panel (c) and 
(d) shows the velocity-time series and wavelet power map of Mg~{\sc ii} k2v line. In case of Mg~{\sc ii} k2v, the significant power is 
concentrated around 3.0 minutes (panel d). Similarly, the panels (e) and (f) show the velocity-time series and wavelet power map of 
C~{\sc ii} line. It is evident that the power lies from 2.0 to 4.0 minutes in the TR (panel d). The gray hatched area on each wavelet power 
outlines the cone of influence (COI). Red-dashed lines on each wavelet power maps outline the 95\% significance level.\\
\textbf{\cite{Torr1998} have described the procedure to estimate the significance contours for the wavelet analysis (see also; wave$\_$signif.pro as provided by Christopher Torrence and Gilbert P. Compo of the University of Colorado). The background (theoretical) spectrum is needed to outline the significant locations (i.e., significance level) within the wavelet power, which can be modeled using this equation.} 
%%%%%%%%%%%%%%%%%%%%%%%%%%%%%%%%%%%%%%%%%%%%%%%%%%%%%%%%%%%%%%%%%%%%
\begin{equation}
\label{eq:white_noise}
{P_{k}} = {{1-\alpha^2} \over {1+\alpha^2-2\alpha(2\pi k/N)}}., 
\end{equation}
%%%%%%%%%%%%%%%%%%%%%%%%%%%%%%%%%%%%%%%%%%%%%%%%%%%%%%%%%%%
Here, k=0,1,.,.,N/2 is the frequency index. \textbf{We have taken $\alpha$ = 0 in the present case, which produce the white noise spectrum (one values at each frequency; flat Fourier spectrum). However, the other values of $\alpha$ lead to the red-noise spectrum (i.e., increasing power with decreasing frequency). The variance of Doppler velocity-time series (or multiplication between the variance of Doppler velocity-time series and 1 (P$_{k}$ =1 for white noise)) provides the appropriate theoretical spectrum, which is similar as described in wave$\_$signif.pro by Christopher Torrence and Gilbert P. Compo. The theoretical spectrum is not same for all the Doppler velocity-time series, however, it changes as per the nature of Doppler velocity-time series.} Finally, the theoretical spectrum is multiplied by cut off value of chi-square as per used significance level \textbf{as described by \cite{Torr1998}}. The cutoff value of the
chi-square distribution is estimated using degree-of-freedoms (DOFs) and significance level (95\%). So, this analysis provides the 
significance value, which is used to outline the 95\% significance level on each map (Figure~\ref{fig:wavelet}; red-dashed lines).\\ 
The cross wavelet between two different time-series, which are forming different heights, is important to understand the common power 
inherited in these time-series. The cross wavelet array is defined as the multiple product of wavelet of one series and complex conjugate 
wavelet of another series (\citealt{Torr1998,Bloomfield2004}). The cross wavelet power is the square of absolute magnitude cross wavelet 
array and it highlights those areas which have high common power. In addition, the wavelet coherence is also important to investigate 
the coherent/incoherent oscillations between two different time-series. The cross wavelet power is normalized by the multiplication of 
the power of both series, which is considered as the coherence (\citealt{Torr1998,Bloomfield2004}).\\
%%%%%%%%%%%%%%%%%%%%%%%%%%%%%%%%%%%%%%%%%%%%%%%%%%%%%%%%%%%%%%%%%%%%%%%%%%%%%%%%%%%%%%%
\begin{figure*}
\centering
\mbox{
\includegraphics[trim = 5.5cm 0.5cm 5.5cm 0.0cm, scale=1.2]{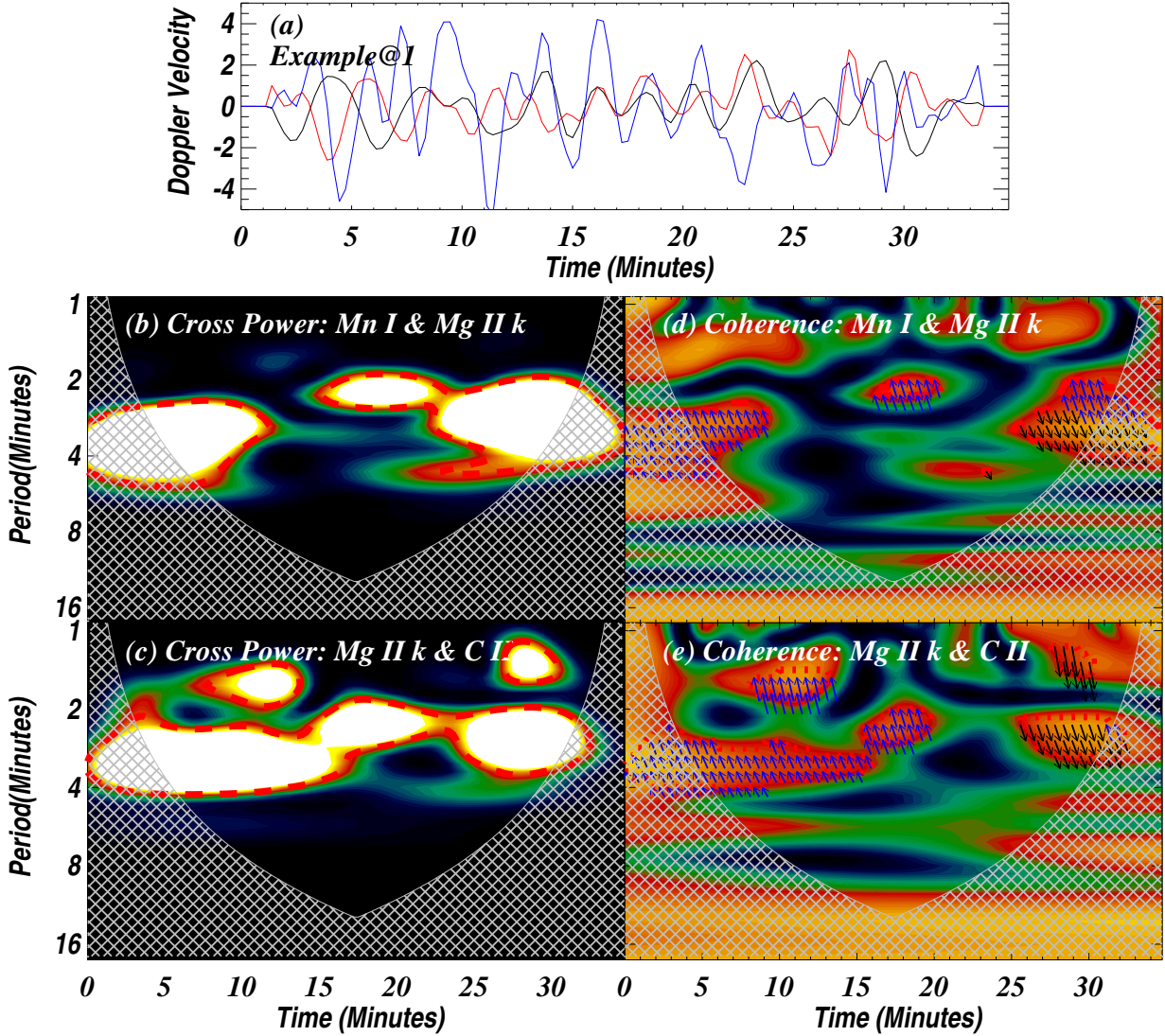}
}
\caption{\small The velocity-time series from Mn~{\sc i} 2801.25~{\AA} (black curve), Mg~{\sc ii} k2v (red curve) and C~{\sc ii} 
(blue curve) (panel a).The cross power between the photosphere and chromosphere (panel b) between the chromosphere and TR (panel c). They 
represent the dominance of cross powers in the period range from 2.0 minutes to 6.0 minutes. Similarly, panel d (panel f) shows the 
coherence map between the photosphere and chromosphere (chromosphere and TR). The over plotted arrows on the wavelet coherence maps 
illustrate the positive (purple arrows) and negative phase lags (black arrow). The waves with period from 2.0 to 4.0 minutes are 
dominated by positive phase lags (panels d $\&$ f), which signifies upward propagation into TR (through chromosphere) from photosphere. 
However, at later times, the waves becomes downwardly propagating waves as indicated by negative phase lags.}
\label{fig:cross_wavelet}
\end{figure*}
%%%%%%%%%%%%%%%%%%%%%%%%%%%%%%%%%%%%%%%%%%%%%%%%%%%%%%%%%%%%%%%%%%%%%%%%%%%%%%%%%%%%%%%
Panel (a) of Figure \ref{fig:cross_wavelet} shows the Doppler velocity-time series of all three lines from one location.   We have estimated 
the cross power between the photosphere (Mn~{\sc i}) and chromosphere (Mg~{\sc ii} k2v) (figure~\ref{fig:cross_wavelet}; panel b).   In case of the photosphere and chromosphere, the cross power is concentrated within the period range of 2.0 to 6.0 minutes (panel b).  However, in case of the chromosphere and TR, most of the common power lies in the period range of 1.5 to 4.0 minutes (Figure~\ref{fig:cross_wavelet}; panel c).   The coherence values can vary between zero and one. The zero value of the coherence represents the complete incoherent oscillations between two time series, while the one value of the coherence corresponds to the perfect coherent oscillations between the two time-series. The cross-spectrum of two time series is necessary to find out significant common power areas in time and frequency domain.   Moreover, the wavelet coherence is also needed to find the co-movements between the two time-series (i.e., two heights in the solar atmosphere).   We also estimate the wavelet coherence between different heights of the solar atmosphere. Panel (d) shows the wavelet coherence, which is drawn using photospheric and chromospheric heights.   Similarly, panel (e) shows the coherence map between chromospheric and TR heights.   The gray hatched area (red dashed) shows the COI (95\% significance level) on each maps.\\ 
We now evaluate phase difference (i.e., difference of phase angles at two different heights) in the time-frequency domains. A cross wavelet 
analysis leads to the complex array in the time-frequency domain, which can be converted into the phase angle using real and imaginary parts
of the complex numbers (\citealt{Torr1998,Bloomfield2004}).   This phase angle (phase lags) is basically the phase difference between two 
different time-series, which are forming at two different heights. We do not use all the phase lags in further analysis. However, we put 
some specific conditions to choose the specific phase lags. Such specific conditions are adopted to increase the reliability of our results.   The adopted specific conditions are as follows: (a) only significant oscillations, which complete more than two cycles at least, are included; (b) we 
exclude the COI area (i.e., figure~\ref{fig:cross_wavelet};gray hatched area); (c) we use only significant common power areas
(95\% significance level; Figure~\ref{fig:cross_wavelet}- red dashed line); (d) the coherence value should be greater than or equal to 0.6. 
The coherence values greater than or equal to 0.6 represent the true co-movements between the two time-series.   So, on the basis of above 
defined criteria, we extract the valid phase difference points in the time-frequency domain.   The valid phase lags are drawn on the wavelet 
coherence maps (panels d and e) by purple color arrows (positive phase-lags) and black color arrows (negative phase-lags). The significant 
coherence between photosphere and chromosphere are dominated by the positive phase lags, which signifies the upward propagation. However, some periods are dominated by negative phase lags, which represent the downwardly propagating waves.   The chromospheric 
and TR heights reveal the significant coherence between the period range of 1.5 to 4.0 minutes, which is also dominated by the positive 
phase lags (panel e).   \textbf{This significant coherence along with positive phase lags} suggest that the photospheric power can successfully reach up to the TR \textbf{within the period regime from 2.0 to 4.0 minutes. Therefore, the cutoff period of $\approx$ 4.0 minutes (i.e., 4.0 mHz) is inferred from this particular pixel with the successful channeling of 3.0 minutes oscillations into TR from photosphere. This observed cutoff period (i.e., 4.0 minutes) is consistent with the classical theory of acoustic wave propagation (\citealt{Lamb1909,Lamb1932,Fleck1993,Routh2014}).}\\
%%%%%%%%%%%%%%%%%%%%%%%%%%%%%%%%%%%%%%%%%%%%%%%%%%%%%%%%%%%%%%%%%%%%%%%%%%%%%%%%%%%%%%%%%%%%%%%%%%%%%%%%%%%%%%%%%%%%%%%%%%%%%
\begin{figure*}
\centering
\mbox{
\includegraphics[trim = 4.5cm 0.5cm 4.5cm 2.0cm, scale=1.1]{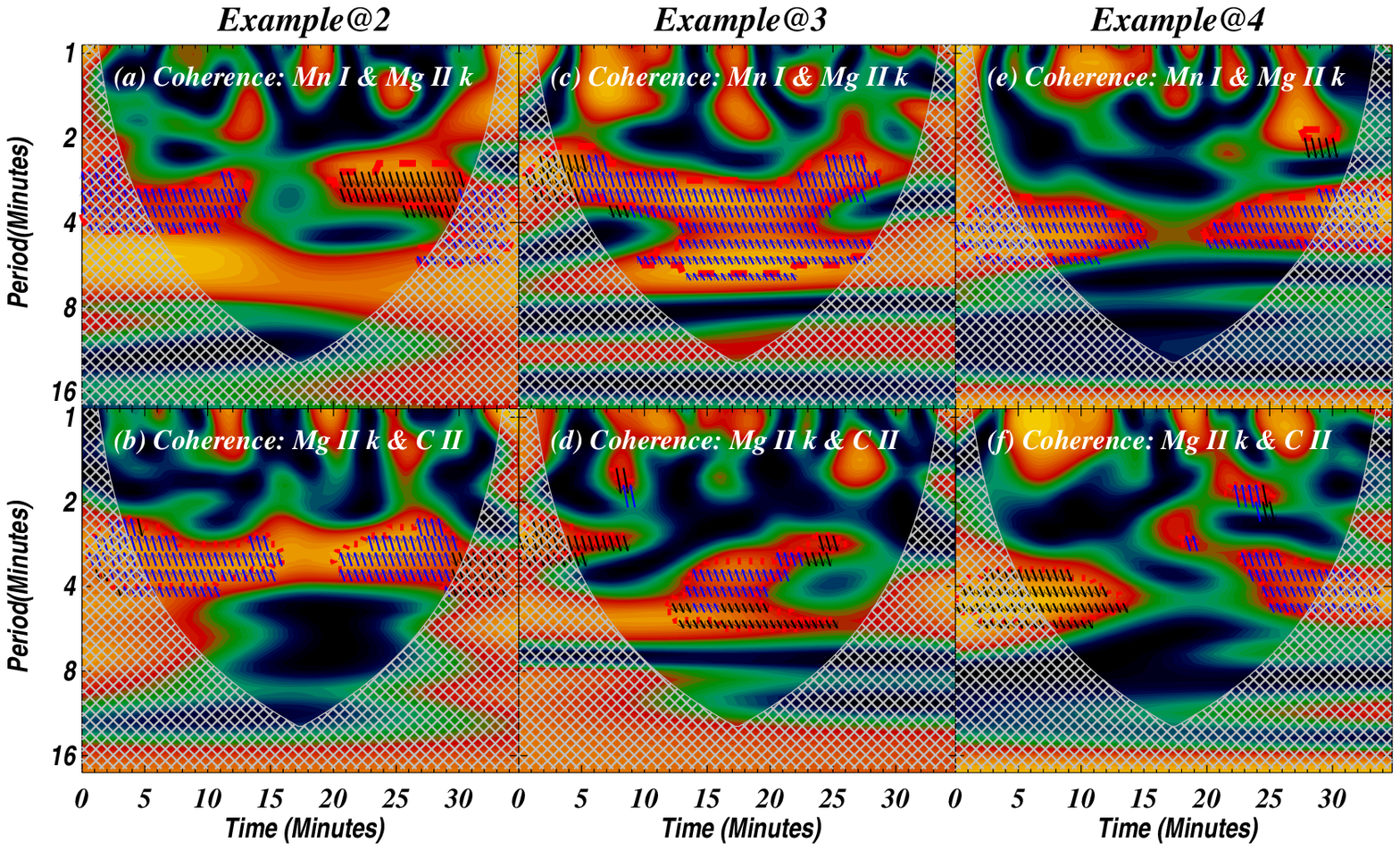}
}
\caption{\small The phase lags along with coherence within the time-period domain using three different locations to show their variations. 
Exapmle 2: the significant coherence is located in two different patches (0-15 minutes $\&$ 20-35 minutes). The first patch is occupied by 
the positive phase lags up to the TR from the photosphere (panel a $\&$ b). The second patch (20-35 minutes) shows negative phase lags between
the chromosphere and photosphere (panel a). However, the power from the chromosphere within the same time and period range reaches into the 
TR (panel b). Example 3: the significant  coherence is located in the period range from 2.0 to 6.0 minutes. All the waves within this period 
range reaches up to the chromosphere from the photosphere (panel c). However, some higher periods (greater than 4.0 minutes) corresponds to
downward propagation (panel d). Example 4: two different time patches (i.e., 0-15 $\&$ 18-35 minutes) are present with the significant 
coherence within the period regime from 3.0 to 6.0 minutes. All the waves within this period range propagate into the chromosphere from the 
photosphere. However, only second patch (18-35 minutes) exhibit the further propagation of power from the chromosphere up to the TR.}
\label{fig:example_propagate}
\end{figure*}
%%%%%%%%%%%%%%%%%%%%%%%%%%%%%%%%%%%%%%%%%%%%%%%%%%%%%%%%%%%%%%%%%%%%%%%%%%%%%%%%%%%%%%%%%%%%%%%%%%%%%%%%%%%%%%%%%%%%%%%%%%%%%%
However, not all the locations show the \textbf{similar pattern of the propagation of waves} through the different layers of the solar atmosphere.   So, we presented some other examples to demonstrate the different behaviors of the wave propagation up to the TR \textbf{from photosphere (Figure~\ref{fig:example_propagate})}.  In example 2, the waves (within the period range from 2.5 to 4.0 minutes) successfully propagate 
upward up to the TR in the time range from 0 to 15 minutes (Figure~\ref{fig:example_propagate}; panel a). However, in the time regime 
of 20.0 to 30.0 minutes, the waves are propagating downward from the chromosphere. Interestingly, the waves are propagating upward from the chromosphere into the TR in the time regime of 20.0 to 30.0 minutes (Figure~\ref{fig:example_propagate}; panel b). \textbf{Therefore, this particular location also follows the wave propagation theory (i.e., cutoff period of 4.0 minutes; successful propagation of 3.0 minutes into TR from photosphere)}. \textbf{On the other hand,} example 3 shows the successful wave propagation \textbf{within the broad range of periods (i.e, 2.0 minutes 6.0 minutes)} from the photosphere up to the chromosphere (Figure~\ref{fig:example_propagate}; panel c). However, the waves propagate upward \textbf{into TR from chromosphere} within a small period range (i.e., 3.0 to 4.0 minutes) and for other periods downward propagation can be inferred (Figure~\ref{fig:example_propagate}; panel d). Similarly, example 4 shows the propagation of waves from the photosphere up to the chromosphere within the \textbf{almost similar} period range (3.0-6.0 minutes) \textbf{for two different} time ranges (0-15 and 15-30 minutes). In one time-regime (0-15 minutes), waves show the downward propagation between the chromosphere and TR. However, in another time regime (15-30 minutes), the waves propagate up to the TR from chromosphere (figure~\ref{fig:example_propagate}; panel f). These examples clearly exhibit the complex behavior of the wave propagation into the TR from the photosphere. \textbf{Interestingly, the successful propagation of waves with the period greater than the cutoff period (i.e., greater than 4.0 minutes; specially 5.0 minutes) into TR from photosphere emerge as the most noticeable feature from this complex behavior. The propagation of 5.0 minutes into TR from photosphere is not allowed by the wave propagation theory. However, the presence of magnetic field can allow to propagate longer periods as reported by \cite{Hegg2011}. Similarly, \cite{DePon2003} and \cite{DePon2005} have also reported that 5.0 minutes reaches into TR from the photosphere within the regime of high magnetic field (i.e., plage region).} \\     
%%%%%%%%%%%%%%%%%%%%%%%%%%%%%%%%%%%%%%%%%%%%%%%%%%%%%%%%%%%%%%%%%%%%%%%%%%%%%%%%%%%%%%%%%%%%%%%%%%%%%%%%%%%%%%%%%%%%%%%%%%%%%%%%%%%%%%%%%%%%%%%%%%%
\begin{figure*}
\centering
\mbox{
\includegraphics[trim = 4.5cm 1.0cm 4.5cm 0.0cm, scale=1.1]{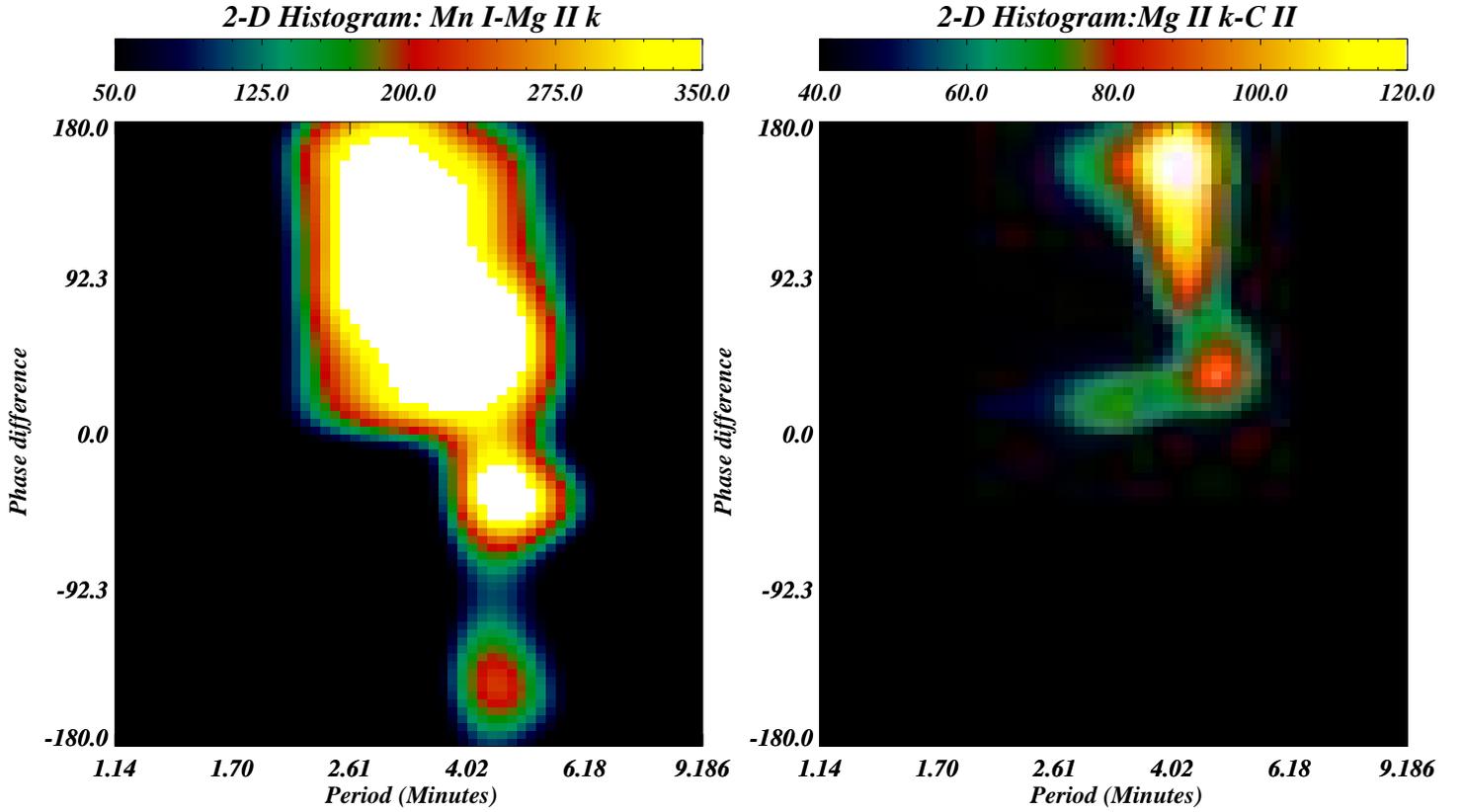}
}
\caption{\small The histogram of phase lags (i.e., period vs phase lags) between the photosphere and chromosphere (left-panel) 
and between the chromosphere and TR (right-panel). The photosphere and chromosphere: a significant histogram density with positive phase 
lags lies within the regime of 3-minute oscillations (left-panel). The regime of 5-minute oscillations are dominated by negative phase lags
that signifies the downward propagation. However, there are locations that show the upward propagation of 5 minutes oscillations too. The 
chromosphere and TR: the oscillation with their period from 2.5 up to 6.0 minutes successfully propagate from the chromosphere into the TR.}
\label{fig:histogram}
\end{figure*}
%%%%%%%%%%%%%%%%%%%%%%%%%%%%%%%%%%%%%%%%%%%%%%%%%%%%%%%%%%%%%%%%%%%%%%%%%%%%%%%%%%%%%%%%%%%%%%%%%%%%%%%%%%%%%%%%%%%%%%%%%%%%%%%%%%%%%%%%%%%%%%%%%%%
\textbf{Using the above described process}, we have collected all the valid phase lags from all the locations to study the statistical behavior of wave propagation. For the statistical behavior, we choose only those regions in the time-frequency domain that show the coherent oscillations from the photosphere up to the TR (i.e., coherent oscillation between the photosphere and chromosphere and then the similar oscillations between the 
chromosphere and TR). Using this procedure, we have produced 2-D histogram of the phase lags between photosphere and chromosphere (Figure~\ref{fig:histogram}; left-panel). In this 2-D histogram, we have shown the distribution of phase lags density with period.   The histogram reveals that the phase lags are distributed within the period range from 2.0 to 6.0 minutes. The highest histogram density with positive phase lags (upward propagation) is concentrated around 3-minute oscillations.   On the contrary, the histogram density is significant around the regime of 5 minutes oscillations, which is dominated by the negative phase lags (downward propagation). However, there is also some significant histogram density with the positive phase lags associated with 5-minute oscillations. In addition, some locations also show the negative phase lags in the regime of 3.0 minutes oscillations too. In the similar fashion, the histogram of phase lags between the chromosphere and TR are also produced (Figure~\ref{fig:histogram}; right-panel). In case of wave propagation between chromosphere and TR, it is clearly visible that most of the spectral density is concentrated in the period ranging from 2.0 to 6.0 minutes. \textbf{The propagation of 5.0 minutes into the TR from photosphere at least at some locations can be inferred from the statistical behavior of phase lags, which may be the result of the presence of magnetic field.}\\
%%%%%%%%%%%%%%%%%%%%%%%%%%%%%%%%%%%%%%%%%%%%%%%%%%%%%%%%%%%%%%%%%%%%%%%%%%%%%%%%%%%%%%%%%%%%%%%%%%%%%%%%%%%%%%%%%%%%%%%%%%%%%%
\begin{figure*}
\centering
\mbox{
\includegraphics[trim = 4.5cm 1.0cm 4.5cm 3.0cm, scale=1.1]{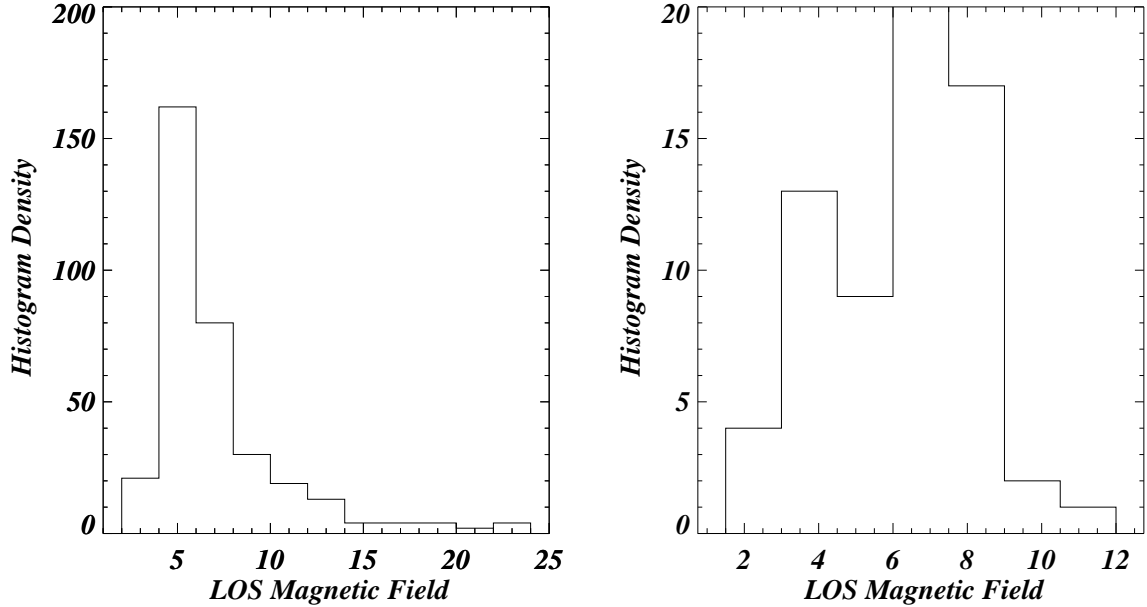}
}
\caption{\small Statistical distribution of LOS photospheric magnetic field for short periods (3.0 minutes; left-panel) and for long periods (5.0 minutes; right-panel). Distribution of LOS photospheric magnetic field for short periods peaks around 5 Gauss while it peaks around 8 Gauss for longer periods. The short period distribution has long tail, which lies beyond 20 Gauss.}
\label{fig:histogram_mag}
\end{figure*}
%%%%%%%%%%%%%%%%%%%%%%%%%%%%%%%%%%%%%%%%%%%%%%%%%%%%%%%%%%%%%%%%%%%%%%%%%%%%%%%%%%%%%%%%%%%%%%%%%%%%%%%%%%%%%%%%
\textbf{The photospheric magnetic field is investigated to understand its effect on wave propagation. HMI/SDO observations are utilized to extract the line-of-sight (LOS) photospheric magnetic field in the vicinity of slit. Now, we have divided the locations into two different classes: (1) locations which allow successful propagation of short periods (3 minutes) and (2) locations which allow successful propagation of longer periods (5 minutes). We have selected LOS photospheric magnetic field for both classes, which are shown in Figure~\ref{fig:histogram_mag}. The left panel of Figure~\ref{fig:histogram_mag} shows the statistical distribution of LOS photospheric magnetic field for the shorter periods (3 minutes). However, right panel of Figure~\ref{fig:histogram_mag} shows the statistical distribution of LOS photospheric magnetic field for the longer periods (5 minutes). The peak of histogram lies around 5 Gauss for short periods (left-panel) while this peak shifts towards slightly higher values of LOS magnetic filed (around 8 Gauss) for longer periods (right-panel). It should be noted that high LOS magnetic field is also present for short periods (long tail of the histogram; left-panel). So, the LOS photospheric magnetic field distribution supports that longer period lies in the regime of high magnetic field. However, this claim is not very convincing because both statistical distribution of LOS photospheric magnetic field are not very different from each other.}
% -----------------------------------------------------------------------------------------
\section{Discussion and conclusions}\label{sec:Summary}
% -----------------------------------------------------------------------------------------
In this work, we have studied the propagation of waves through the different layers of the solar atmosphere using IRIS multi-line 
observations (Mn~{\sc i} 2803.25~{\AA} - the photosphere, Mg~{\sc ii} k2v - middle chromosphere, and C~{\sc ii} - TR).   The
 wavelet analysis predicts the significant coherent oscillations between the photosphere and chromosphere.   In addition, the phase
 lags indicate that most of the photospheric wave power (around 70\% locations) propagates vertically upward up to the chromosphere
 in the regime of high frequencies (i.e., wave periods ranging from 1.8 to 4.0 minutes).   On the contrary, the lower frequencies 
 (5-minutes oscillations) shows the downward propagation (negative phase lags) between the photosphere and chromosphere.   However, 
 some locations also reveal the propagation of lower frequencies into chromosphere from the photosphere.   Majority of the locations 
 do not allow the vertical propagation of waves beyond 4 minutes oscillations, therefore, we can say that the cutoff period between 
 the photosphere and chromosphere is approximately 4 mHz.   In case of the chromosphere and TR, most of the power 
 (in significant period range from 2.6 to 6.0 minutes) reaches up to the TR from the chromosphere.   It must be noted that not all the 
 locations (i.e., 70\% locations; vertical propagation between photosphere and chromosphere) exhibit significant vertical 
 propagation from the chromosphere into the TR.   However, only 45 \% locations out of those 70\% have coherent oscillations 
 between the chromosphere and TR with positive phase lags.   It is also evident that 5-minutes oscillations corresponds to most 
 dominating propagating period between the chromosphere and TR. \\
   
The region, which is used to study the wave propagation, is practically magnetic field-free. \textbf{IN regions, which lie along the slit, are used to be very weak or free magnetic field regions. The distribution of LOS photospheric magnetic field supports the very weak nature of magnetic field within the vicinity of IRIS slit. Accordingly, we have considered the used region as magnetic field-free regions.} Therefore, the considered waves are essentially acoustic.   The propagation of acoustic waves was extensively studied by using observations as well as numerical 
simulations (\citealt{Lites1979, Lites1982, Carlsson1992, Carlsson1997, Wik2000,Judge2001,Murawski2016,Muraw2016,Kont2016,Wisi2016}). 
The 3-minute oscillations (i.e., 5.5 mHz) have very low power when compared to the 5-minutes oscillations (3.3 mHz) at the solar 
photosphere. However, 3 minute oscillations are dominant in the chromosphere, where the 5-minute oscillations become evanescent
(\citealt{Lites1982,Cheng1996,Carlsson1997,Murawski2016}). \cite{Centeno2006, Centeno2009} reported that the chromospheric 3-minute
oscillations are correlated with the photospheric 3 minute oscillations in the sunspot umbra. They proposed that the chromospheric 
3-minute oscillations originate from the photospheric power.   The acoustic wave propagation theory (i.e., for isothermal atmosphere)
predicts free propagation of 3-minute oscillations up to the TR from the solar photosphere (\citealt{Lamb1909,Lamb1910}).  
Similar results were obtained by more recent theoretical work (e.g, \citealt{Fawzy2012,Routh2014}) in which more 
realistic (\citealt{Vernazza1981}) solar atmosphere models were adopted. \\

It is noteworthy that different types of dynamical events in the real solar atmosphere are continuously happening, which can 
modify/change/destroy the photospheric power as it travels through the different layers of the solar atmosphere.   In the present analysis, 
most of locations (around 70\%) predicts the successful upward propagation of acoustic waves with higher frequencies into the 
chromosphere from the photosphere.   The rest of the positions (i.e., 30\%), which do not fit the correlation (e.g., coherence, 
positive phase lags, etc.) between photosphere and chromosphere, may be modified due to the presence of complex/dynamical 
events in the atmosphere between photosphere (i.e., Mn~{\sc i} spectral line) and chromosphere (Mg~{\sc ii} k2v spectral line).\\
 
\textbf{In active-region plage,} \citealt{DePon2003} have reported that only 50\% locations showed \textbf{successful propagation of 5.0 minutes into the TR from photosphere. The propagation of longer period into the upper atmosphere from photosphere is the result of magnetic field (inclinations). However, the present analysis shows that} 5-minute oscillations, between the photosphere and chromosphere, are dominated (\textbf{most of the locations}) by the negative phase lags, which supports the evanescent nature of 5-minute oscillations as predicted by the linear wave theory. \textbf{The unsucessful propagation of 5 minute into chromosphere (or cutoff period $\approx$ 4.0 as per linear theory) predicts no influence from magnetic field (i.e., no magnetic field or may be the vertical magnetic field). Interestingly,} the successful propagation 
of 5-minute oscillations is also present \textbf{at some locations. } The photospheric 5 minute power can reach up to the TR/corona in the regime of magnetic field (plage, flux-tubes; \citealt{DePon2003,DePon2005}). \textbf{Although, the used region} has very weak magnetic field, yet, some bright patches are visible in the chromosphere/TR (cf., Figure~\ref{fig:ref_image}), which are associated with the magnetic field. \textbf{In addition, we have also shown that the LOS photospheric magnetic field is slightly higher at those locations where 5-minute is reaching upto the TR from photosphere.} Therefore, the propagation of 5.0-minutes \textbf{into TR is associated with magnetic field}. In case of correlation between the 
chromosphere and TR, oscillations with their period range between 2.5 and 6.0 minutes successfully propagate from the chromosphere to the TR.   However, only 45 \% locations (out of 70\%) show the correlation between the chromosphere and TR. It should be noted that most of power is transferred into the TR from the chromosphere through the 5-minute oscillations.

The focus of this paper is on different aspects of the acoustic wave propagation than that investigated by \citealt{Wisi2016}.  Nevertheless,
there is consistence between the results of both papers.  First of all, both papers clearly demonstrate the existence of the acoustic cutoff
frequency in the solar atmosphere, and its important role in setting up the wave propagation conditions.  Second, both papers show similar
ranges for propagating and non-propagating acoustic waves.  Moreover, the results of this paper clearly emphasize \textbf{on the nature of 3 and 5-minutes oscillations and their propagation/non-propagation in the TR from photosphere.}
%the role of photospheric oscillations in both driving the 3-min chromospheric oscillation, and transferring the power of 5-min oscillations to the solarand transition-region.  
The presented results also show how efficient is wave reflection in the solar atmosphere by establishing percentage of propagated to transferred waves.   Finally, our results have significant implications on theoretical work as they indicate that propagating acoustic waves may drive the 3-min chromospheric oscillations (\citealt{Fleck1991}, Sutm1998) but they also leave a room for possible existence of a wave cavity in the solar chomosphere as suggested by recent numerical work of Kar2018.   More theoretical work is required in order to resolve these important problems related to the origin of atmospheric oscillations in the solar atmosphere.   

We conclude that the photospheric power can propagates up to the TR in the form of 3-minute \textbf{as well as 5 minute (at least at some locations)} oscillations.  Therefore, the 3 minutes chromospheric/TR oscillations in the internetwork regions \textbf{(regions without the influence of magnetic field)} are directly powered by the photospheric oscillations. \textbf{In addition, a few locations also show the propagation of 5-minute into TR from photosphere, which might be the result of magnetic field. A} spectrum of wave periods (i.e., from 2.5 minutes to 6.0 minutes) in the TR is directly associated with the corresponding chromospheric spectrum of wave periods.   Therefore, the low chromospheric frequencies (i.e., in the regime of 5-minutes oscillations) are sources of the corresponding low frequencies of the TR.   
   
\section*{Acknowledgements}
PK's and KM’s work was done in the framework of the project from the National Science Centre, Poland (NCN), Grant No. 2014/15/B/ST9/00106.  ZEM acknowledges partial support of his work by the Alexander von Humboldt Foundation.   IRIS is a NASA small explorer 
mission developed and operated by LMSAL with mission operations executed at NASA Ames Research center and major contributions 
to downlink communications funded by ESA and the Norwegian Space Centre. AKS and BND acknowledge the RESPOND-ISRO project, 
while AKS acknowledges the SERB-DST young scientist project grant.

\end{document}